\documentclass[aps,prl,reprint,amsfonts,amsmath,amssymb,showpacs,groupedaddress,superscriptaddress]{revtex4-2}
\usepackage[colorlinks,urlcolor=blue,linkcolor=blue,anchorcolor=blue,citecolor=blue,bookmarks]{hyperref}
\usepackage{graphicx}
\usepackage{bm}
\usepackage{color}
\usepackage{subfigure}

\begin{document}

\title{Nature of the 1/3 Magnetization Plateau in Spin-1/2 Kagome Antiferromagnets}
\author{Li-Wei He}
\affiliation{National Laboratory of Solid State Microstructures and Department of Physics, Nanjing University, Nanjing 210093, China}
\author{Xinzhe Wang}
\affiliation{National Laboratory of Solid State Microstructures and Department of Physics, Nanjing University, Nanjing 210093, China}
\author{Shun-Li Yu}
\email{slyu@nju.edu.cn}
\affiliation{National Laboratory of Solid State Microstructures and Department of Physics, Nanjing University, Nanjing 210093, China}
\affiliation{Collaborative Innovation Center of Advanced Microstructures, Nanjing University, Nanjing 210093, China}
\author{Jian-Xin Li}
\email{jxli@nju.edu.cn}
\affiliation{National Laboratory of Solid State Microstructures and Department of Physics, Nanjing University, Nanjing 210093, China}
\affiliation{Collaborative Innovation Center of Advanced Microstructures, Nanjing University, Nanjing 210093, China}

\date{\today}

\begin{abstract}
We investigate the origin of the 1/3 magnetization plateau in the $S=1/2$ kagome antiferromagnetic Heisenberg model using the variational Monte Carlo and exact diagonalization methods, to account for the recent experimental observations in YCu$_3$(OH)$_{6+x}$Br$_{3-x}$ and YCu$_3$(OD)$_{6+x}$Br$_{3-x}$. We identify three degenerate valence-bond-solid (VBS) states forming a $\sqrt{3} \times \sqrt{3}$ unit cell. These states exhibit David-star patterns in the spin moment distribution with only two fractional values $-1/3$ and $2/3$, and are related through translational transformations. While the spin correlations in these VBS states are found to be short-range, resembling a quantum spin liquid, we show that they have a vanishing topological entanglement entropy and thus are topologically trivial many-body states. Our theoretical results provide strong evidence that the 1/3 magnetization plateau observed in recent experiments arises from these $\sqrt{3} \times \sqrt{3}$  VBS states with fractional spin moments.
\end{abstract}

\maketitle

\begin{figure}
    \centering
    \subfigure{
    \begin{minipage}{0.3\linewidth}
    \centering
    \includegraphics[width = 1.0\linewidth]{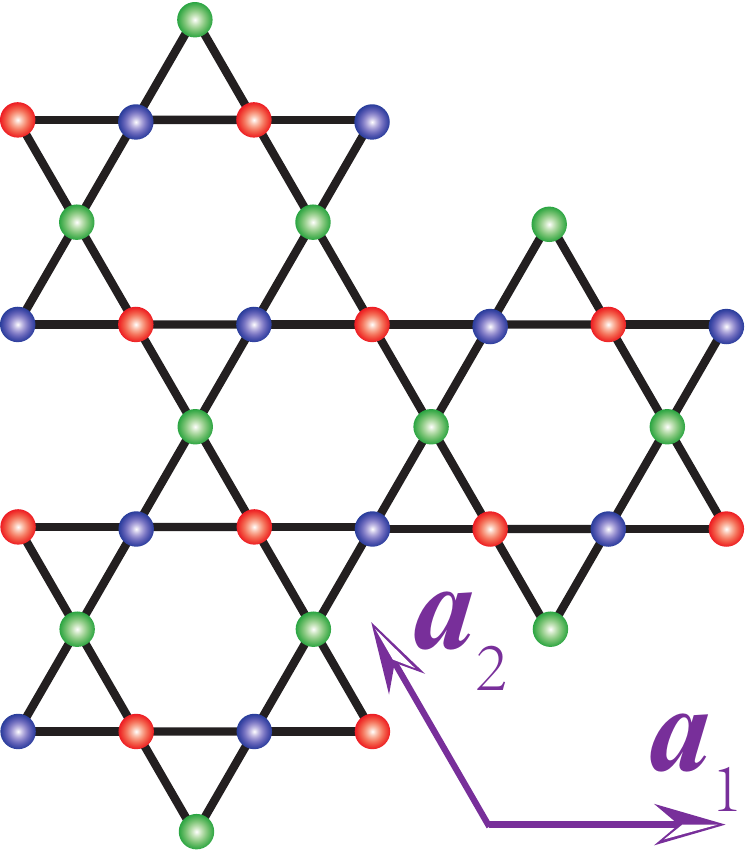}
    \put(-75,80){(a)}
    \label{fig:kagome_lattice}
    \end{minipage}
    }%
    \subfigure{
    \begin{minipage}{0.67\linewidth}
    \centering
    \includegraphics[width = \linewidth]{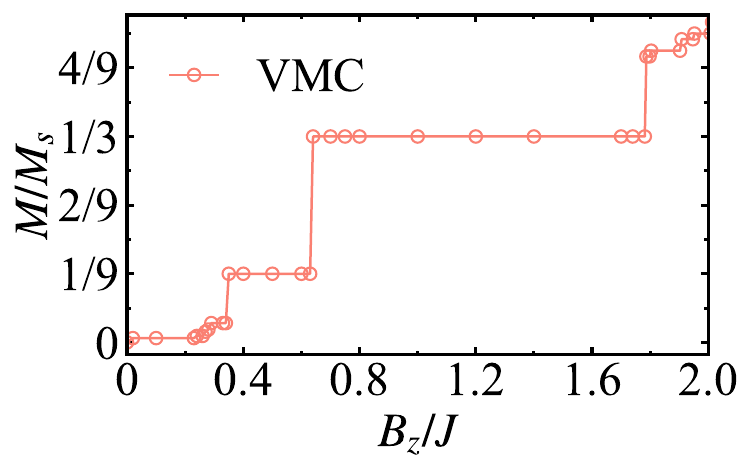}
    \put(-168,88){(b)}
    \label{fig:m_vs_bz_vmc_ed}
    \end{minipage}
    }

    \caption{(a) Kagome lattice. The three different colors represent the three sublattices. The purple arrows indicate the basis vectors $\boldsymbol{a}_{1} = (1, 0)$ and $\boldsymbol{a}_{2} = (-1/2, \sqrt{3}/2)$ of its primitive cell. (b) displays the average magnetization $M/M_s$ as a function of the magnetic field $B_{z}/J$, obtained using the VMC method for a large system with at least $N = 12 \times 12 \times 3 = 432$ sites.}
    \label{fig:kagome_magnet_cuuve}
\end{figure}

{\it Introduction.} The kagome lattice, a two-dimensional network of corner-sharing triangles arranged in a hexagonal motif [as shown in Fig.~\ref{fig:kagome_lattice}], has emerged as one of the most versatile platforms in condensed matter physics. Its unique geometry gives rise to a variety of exotic quantum phenomena, including novel quantum magnetism~\cite{PhysRevLett.101.026405,PhysRevB.83.214406,nat.555.638,nat.562.91}, unconventional superconductivity~\cite{PhysRevB.79.214502,PhysRevB.85.144402,PhysRevLett.125.247002, nat.599.222,nat.phys.18.137, cpb.31.097405, nat.612.647,nat.604.59, nsr.10.nwac199,nat.rev.phys.5.635,nat.rev.mat.9.420,PhysRevLett.134.106401}, charge density wave orders~\cite{nat.phys.18.137,nat.rev.phys.5.635,PhysRevB.85.144402,nat.604.59, nat.611.682,PhysRevLett.134.086902,PhysRevLett.134.096401} and topological quantum states~\cite{PhysRevB.80.113102,PhysRevB.80.193304,PhysRevB.82.075125,PhysRevLett.115.186802,PhysRevB.81.235115,PhysRevLett.106.236802,PhysRevB.83.165118,science.365.1282}. Of particular interest is its potential to realize quantum spin liquids (QSLs)~\cite{PhysRevLett.98.117205,PhysRevLett.101.117203,science.332.1173,PhysRevLett.109.067201,PhysRevB.87.060405,PhysRevLett.118.137202,PhysRevX.7.031020,PhysRevB.95.235107,sci.bull.63.1545,sci.adv.4.eaat5535,PhysRevB.110.035131}, owing to the remarkable combination of strong geometric frustration and a low coordination number. Theoretically, while there is still no consensus on the precise nature of the ground state of the spin-$1/2$ kagome antiferromagnetic Heisenberg model, many studies suggest that the ground state of the system could be a quantum spin liquid. On the experimental front, kagome antiferromagnets such as herbertsmithite~\cite{PhysRevLett.98.107204,nat.492.406,science.350.655,nat.phys.16.469}, Zn-barlowite~\cite{cpl.34.077502,PhysRevB.98.155127,cpl.37.107503,npj.quantum.mater.5.74,nat.commun.12.3048}, and YCu$_3$(OH)$_{6+x}$Br$_{3-x}$~\cite{PhysRevB.105.L121109,PhysRevB.110.085146} have shown great promise as QSL candidates.

Recent breakthroughs in kagome magnetism have further highlighted its unique role. In particular, experiments in two kagome antiferromagnets, YCu$_3$(OH)$_{6+x}$Br$_{3-x}$ and YCu$_3$(OD)$_{6+x}$Br$_{3-x}$~\cite{nat.phys.20.1097,nat.phys.20.435,PhysRevLett.132.226701,arxiv.2409.05600,pnas.2421390122}, have revealed a striking sequence of magnetization plateaus under magnetic fields. At low fields, a 1/9 magnetization plateau phase emerges, while a 1/3 magnetization plateau occurs at higher fields. The 1/9 plateau has been interpreted as either a $\mathbb{Z}_{3}$ QSL, as suggested by variational Monte Carlo (VMC)~\cite{PhysRevLett.133.096501} and density matrix renormalization group (DMRG) simulations of a $S=1/2$ kagome antiferromagnet~\cite{nat.commun.4.2287}, or a valence-bond-solid (VBS) state, as proposed by tensor network calculations~\cite{PhysRevB.93.060407,PhysRevB.107.L220401}. The 1/3 magnetization plateau in a kagome antiferromagnet Heisenberg model has also been found in DMRG simulations~\cite{nat.commun.4.2287} and is suggested to the origin from a VBS with a configuration that each triangular unit consisting of one fully polarized spin with scaled moment $M_{i}/M_{s} = 1$ ($M_{s} = 1/2$ is the saturated magnetization for spin-1/2 system) and two spins forming a singlet.

%For the 1/3 magnetization plateau, although theoretical consensus suggests its origin as a valence bond solid, its exact properties still require further detailed study.

In this Letter, we identify theoretically three degenerate VBS states with a unit-cell size of $\sqrt{3} \times \sqrt{3}$ within the 1/3 magnetization plateau phase of a $S=1/2$ kagome antiferromagnetic Heisenberg model, by use of the VMC and exact diagonalization (ED) methods. The magnetization distributions of these VBS states differ fundamentally from that proposed in Ref.~\onlinecite{nat.commun.4.2287}, in that each triangular unit consists of a fractional $M_{i}/M_{s} = -1/3$ spin moment and two fractional 2/3 moments. This feature of fractional spin moments demonstrates that the 1/3 magnetization plateau phase contributed by these VBS states can't be described by semiclassic linear spin-wave theory, which treats the spin as a classic three-dimensional vector. The relevance of the VBS states has been checked by calculating the overlap between their wave functions and those obtained by ED method and the result shows that the overlap values are $\gtrsim 60\%$ compared to a value  $33\%$ for the state envisaged in DMRG simulations~\cite{nat.commun.4.2287}. Thus,  we ascribe the 1/3 magnetization plateau observed in recent experiments ~\cite{nat.phys.20.435,PhysRevLett.132.226701,arxiv.2409.05600,pnas.2421390122} to the origin from the VBS states with two fractional moments. We further find that these VBS states exhibit short-range spin-spin correlation, which is quite similar to that in the $\mathbb{Z}_{3}$ QSL~\cite{PhysRevLett.133.096501}. However, they have a zero topological entanglement entropy, indicating the lack of an intrinsic topological order which is intrinsically different from a gapped QSL.

%explore the nature of the 1/3 magnetization plateau by combining VMC and exact diagonalization (ED) methods. We identify three degenerate VBS states with a unit-cell size of $\sqrt{3} \times \sqrt{3}$ within the 1/3 magnetization plateau phase. Although their magnetization distributions are related by the translation transformations $T_{1,2}: \boldsymbol{r} \to \boldsymbol{r} + \boldsymbol{a}_{1,2}$, these transformations are not symmetric because their unit cells are three times larger than the primitive cell of the kagome lattice. The magnetization distributions of these VBS states differ significantly from those previously proposed in Ref.~\onlinecite{nat.commun.4.2287}. Similar to QSLs, the spin correlations of the VBS states in the 1/3 magnetization plateau phase are short-range. However, unlike the QSL in the 1/9 magnetization phase, the zero topological entanglement entropy indicates that these VBS states lack intrinsic topological order.

{\it Model and Method.} The Hamiltonian of the spin-$1/2$ kagome antiferromagnet with a magnetic field is given by
\begin{equation}
    H = J\sum_{\langle ij \rangle} \boldsymbol{S}_i \cdot \boldsymbol{S}_j - B_z \sum_i S_i^z,
    \label{eq:model}
\end{equation}
where $\langle ij \rangle$ denotes the sum over nearest-neighbor bonds, $\boldsymbol{S}_i$ represents the spin-$1/2$ operator at site $i$ (with $S_i^z$ representing its $z$-component), $J$ is the exchange interaction, and $B_{z}$ is the magnitude of the external magnetic field. For finite $B_{z}$, the total $S^{z}$ remains conserved.

In the standard VMC process, we introduce the fermionic doublet to represent the spin operator, $S^{\alpha}_{i} = \frac{1}{2} \psi_{i}^{\dagger} \sigma_{\alpha} \psi_{i}$, where $\alpha \in \{x, y, z\}$, $\psi = (c_{i\uparrow}, c_{i\downarrow})^{T}$ and $c_{i\uparrow(\uparrow)}$ are the annihilation operators of the spin-up (down) spinons, respectively, and $\sigma_{\alpha}$ are the Pauli matrices. For spin-1/2 system, the single-occupancy constrain $N_{i} = n_{i\uparrow} + n_{i\downarrow} = \psi_{i}^{\dagger} \psi_{i} = 1$ is necessary. We then decouple the model (\ref{eq:model}) into its non-interacting mean-field form using the above representation:
\begin{equation}
    H_{\mathrm{mf}} = \sum_{\langle ij \rangle} t_{ij} \psi_{i}^{\dagger} \psi_{j} + \mathrm{h.c.} - \mu \sum_{i} \psi_{i}^{\dagger} \sigma_{z} \psi_{i},
    \label{eq:h_mf}
\end{equation}
where $t_{ij}$ are the spinon hopping terms between the NN site $i$ and $j$, $\mu$ is the chemical potential modulated by the field $B_{z}$ and determines the magnetization. The system size used in the VMC method is always finite, and the magnetization should increase gradually in the form of stpdf with the field for a system that conserves total $S^{z}$ conservation. The ground state of $H_{\mathrm{mf}}$ can be denoted as $|\mathrm{GS}(t_{ij})_{\mathrm{mf}} \rangle$ as a function of the hopping parameters $t_{ij}$. After applying the Gutzwiller projection $P_{G} = \prod_{i} (1 - n_{i\uparrow}n_{i\downarrow})$, the many-body trial wave function is obtained as $|\Psi(t_{ij})\rangle = P_{G} |\mathrm{GS}(t_{ij})_{\mathrm{mf}} \rangle$. The $t_{ij}$ parameters are treated as variational parameters in the VMC calculation and are determined by minimizing the expectation value $E(t_{ij}) = \langle \Psi(t_{ij}) | H | \Psi(t_{ij})\rangle/\langle \Psi(t_{ij})| \Psi(t_{ij})\rangle$. Our main results are obtained for a system with $L_{1,2} = 12$, which are the lengths along the basis vectors $\boldsymbol{a}_{1,2}$ as shown Fig.~\ref{fig:kagome_lattice}, unless otherwise stated.

\begin{figure}
    \centering
    \subfigure{
    \begin{minipage}{\linewidth}
    \centering
    \includegraphics[width = \linewidth]{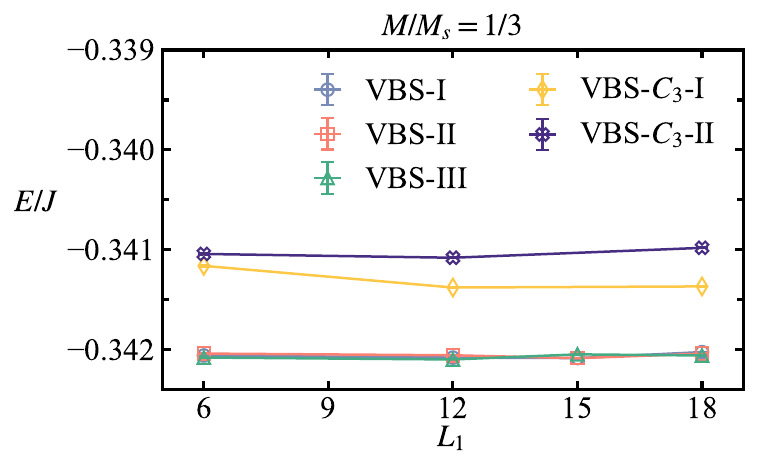}
    \put(-240,130){(a)}
    \label{fig:en_vs_size}
    \end{minipage}
    }
    \subfigure[]{
    \begin{minipage}{0.26\linewidth}
    \centering
    \includegraphics[width = \linewidth]{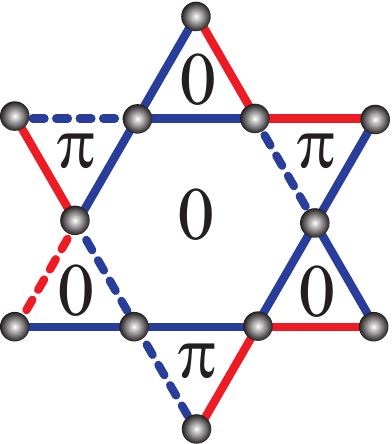}
    \put(-45,76){VBS-I}
    \label{fig:vbs_1}
    \end{minipage}
    }%
    \subfigure[]{
    \begin{minipage}{0.26\linewidth}
    \centering
    \includegraphics[width = \linewidth]{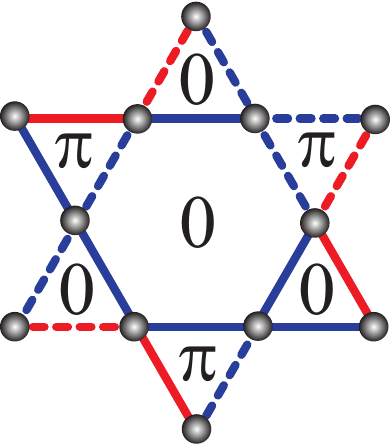}
    \put(-48,76){VBS-II}
    \label{fig:vbs_2}
    \end{minipage}
    }%
    \subfigure[]{
    \begin{minipage}{0.26\linewidth}
    \centering
    \includegraphics[width = \linewidth]{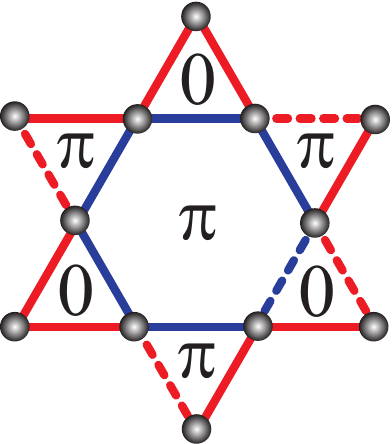}
    \put(-48,76){VBS-III}
    \label{fig:vbs_3}
    \end{minipage}
    }
    \subfigure[]{
    \begin{minipage}{0.26\linewidth}
    \centering
    \includegraphics[width = \linewidth]{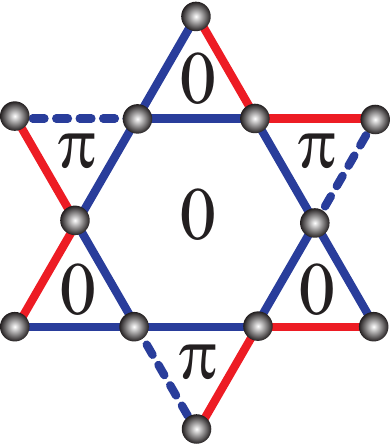}
    \put(-52,76){VBS-$C_{3}$-I}
    \label{fig:vbs_pi_c3}
    \end{minipage}
    }%
    \subfigure[]{
    \begin{minipage}{0.26\linewidth}
    \centering
    \includegraphics[width = \linewidth]{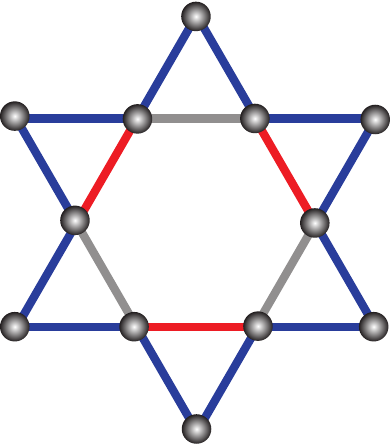}
    \put(-53,76){VBS-$C_{3}$-II}
    \label{fig:vbs_c3}
    \end{minipage}
    }
    \subfigure{
    \begin{minipage}{0.26\linewidth}
    \centering
    \includegraphics[width = \linewidth]{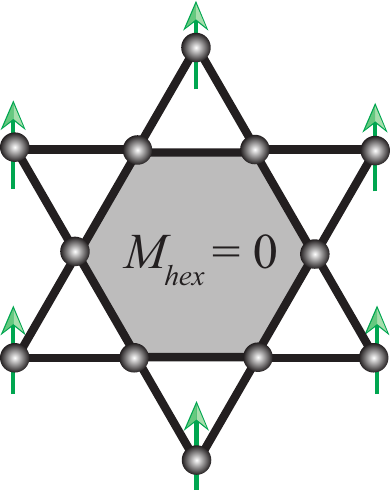}
    \put(-40,-15){(g)}
    \label{fig:vbs_nc}
    \end{minipage}
    }
    \caption{(a) Variational energies per site (excluding the contribution $E(B_{z}) = -B_{z}/6$ from the magnetic field) for several VBS states with 1/3 magnetization are shown as a function of the length $L_{1}$ along vector $\boldsymbol{a}_{1}$ for system sizes of $L_{1} \times 12 \times 3$. (b)-(f) depict the unit-cell patterns of the ``VBS-I",``VBS-II", ``VBS-III", ``VBS-$C_{3}$-I" and ``VBS-$C_{3}$-II" states. Solid (dashed) bonds indicate positive (negative) signs of the spinon hoppings, while red and blue bonds represent different hopping magnitudes. The gray bonds in (f) indicate vanishing hopping terms. (g) shows the magnetization pattern of the VBS state proposed for the 1/3 magnetization plateau in Ref.~\onlinecite{nat.commun.4.2287}. The green arrows represent fully polarized spins, and the magnetization of the shaded hexagon is zero.}
    \label{fig:en_size_dif_vbs}
\end{figure}

{\it Results.} As shown in Fig.~\ref{fig:m_vs_bz_vmc_ed}, a broad 1/9- and a 1/3-magnetization plateau phases are identified using the VMC method, aligning qualitatively with experimental observations~\cite{nat.phys.20.435,PhysRevLett.132.226701,arxiv.2409.05600,pnas.2421390122} and theoretical results~\cite{PhysRevLett.133.096501, nat.commun.4.2287, PhysRevB.93.060407,PhysRevB.107.L220401}. Given the detailed previous VMC studies on the 1/9 magnetization plateau phase~\cite{PhysRevLett.133.096501}, we focus here on the 1/3 magnetization plateau phase. The magnetic field range for the 1/3 plateau is $0.64 \lesssim B_{z}/J \lesssim 1.79$, slightly broader than that obtained via DMRG calculations~\cite{nat.commun.4.2287} due to the difficulty in precisely capturing intermediate states with the VMC method. For the 1/3-magnetization plateau, we identify three energetically favorable VBS states as potential ground states, which are degenerate within numerical error, as shown in Fig.~\ref{fig:en_vs_size}. Furthermore, the overlaps between them are zero, indicating that these VBS states are orthogonal and distinctly different. Their unit cells are three times larger than the primitive cell of the kagome lattice, comprising nine sublattices. In VMC process, we can fix one of the hoppings to be 1 as the reference, then the number of independent hopping terms is 17. Due to the large space of variational parameters, we perform repeated VMC calculations starting from various initial parameters. Ultimately, we identify the three VBS states with hopping patterns displayed in Fig.~\ref{fig:vbs_1}, \ref{fig:vbs_2} and \ref{fig:vbs_3}, respectively, where gauge fluxes and the magnitudes of two types of hopping terms maintain $C_{3}$ rotational symmetry. However, if we artificially adjust the positions of the hopping terms with negative signs, while keeping the flux pattern unchanged to give the David star $C_{3}$ rotational symmetry, this artificial VBS state, shown in Fig.~\ref{fig:vbs_pi_c3}, is not energetically favorable, as seen in Fig.~\ref{fig:en_vs_size}. Additionally, the energy per site of the VBS state named ``VBS-$C_{3}$-II" in Fig.~\ref{fig:en_vs_size} is the highest. Its hopping pattern is derived from the previous work~\cite{PhysRevLett.133.096501}, featuring a $2\pi/3$ flux through each primitive cell of the kagome lattice due to complex hopping terms. We find the optimal gauge fluxes are zero (mod $2\pi$) in all minimal triangles, with three hopping terms nearly vanishing. By eliminating the phase of these hopping terms, we find this artificial state is equivalent to the unmodified one. Therefore, we present the reduced hopping pattern with $C_{3}$ rotational symmetry, as shown in Fig.~\ref{fig:vbs_c3}.

\begin{table}
    \centering
    \caption{Overlaps $\mathcal{O} = |\langle \mathrm{VBS}| \Psi_{\mathrm{ED}}\rangle|^{2}$, where $|\mathrm{VBS}\rangle$ denotes the wave functions of VBS-I, -II, -III states and the VBS-Ref state proposed in Ref.~\cite{nat.commun.4.2287}, respectively, and $|\Psi_{\mathrm{ED}}\rangle$ is the ground-state wave function obtained by ED method in the cluster with $3 \times 3 \times 3 = 27$ sites. We note the optimal variational parameters in the VBS-I, -II and -III states from the system with $12 \times 12 \times 3$ sites.}
    \begin{tabular*}{\linewidth}{@{\extracolsep\fill}l c c c c}
        \hline \hline
        State & VBS-I & VBS-II & VBS-III & VBS-Ref\\
        \hline
        $\mathcal{O}$ & 0.5946  &  0.6880 & 0.5934 & 0.3335\\
        \hline \hline
    \end{tabular*}
    \label{tab:overlap_with_ed}
\end{table}

As a benchmark, we perform an ED calculation on a small cluster with $N = 3 \times 3 \times 3 = 27$, which is the upper limit of our computational capacity. We then calculate the maximal overlaps between the ground state obtained via the ED method and the VBS-I, -II and -III states, as listed in Tab.~\ref{tab:overlap_with_ed}. These states share a significant weight ($\gtrsim 60\%$) of the ED ground states, with minimal differences among the three VBS states. Additionally, we also examine the 1/3-plateau VBS state as defined in Ref.~\onlinecite{nat.commun.4.2287}, where the spins at the six vertices of the David star are fully polarized to 1/2, while the magnetization of the internal hexagon remains zero, as illustrated in Fig.~\ref{fig:vbs_nc}. For the sake of simplicity, we refer to this state as the ``VBS-Ref" state. According to this definition, the representation of the ``VBS-Ref" state in the $3 \times 3 \times 3$ cluster is given by:
\begin{equation}
     | \Psi_{\mathrm{Ref}} \rangle = \prod_{i=1}^{N_{\mathrm{star}}} | \phi_{i} \rangle \prod_{j \in \mathrm{vertex}}| \uparrow_{j} \rangle,
\end{equation}
where $N_{\mathrm{star}} = 3$  is the number of David stars in this cluster, $| \phi_{i} \rangle$ represents the ground state of the hexagon with total $S^{z}=0$ in the $i$-th David star, solved using the ED method, and $|\uparrow_{j} \rangle$ denotes the fully polarized spin state at site $j$. We then calculate the maximal overlap $|\langle \Psi_{\mathrm{Ref}} |\Psi_{\mathrm{ED}} \rangle|^{2}$ and list it in the last column in Tab.~\ref{tab:overlap_with_ed}. This value is nearly half as small as
 those of the three VBS states obtained from the VMC calculations. Based on the ED method benchmark, we believe our VMC results are sufficiently reliable.

\begin{figure}
    \centering
    \subfigure{
    \begin{minipage}{\linewidth}
    \centering
    \includegraphics[width = \linewidth]{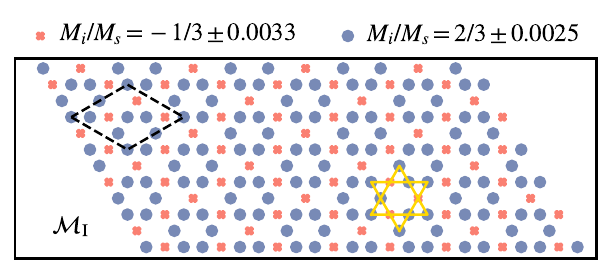}
    \put(-245,93){(a)}
    \label{fig:mz_site_1}
    \end{minipage}
    }
    \subfigure{
    \begin{minipage}{\linewidth}
    \centering
    \includegraphics[width = \linewidth]{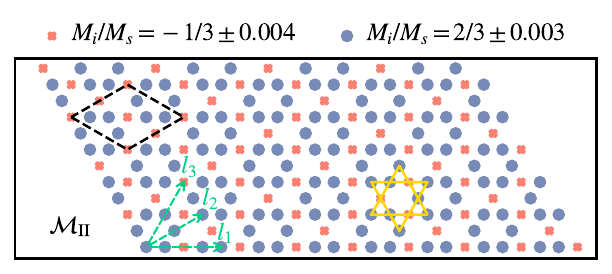}
    \put(-245,93){(b)}
    \label{fig:mz_site_2}
    \end{minipage}
    }
    \subfigure{
    \begin{minipage}{\linewidth}
    \centering
    \includegraphics[width = \linewidth]{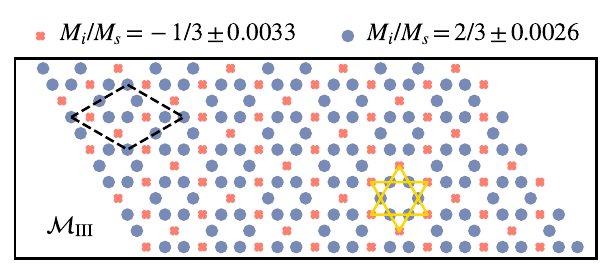}
    \put(-245,93){(c)}
    \label{fig:mz_site_3}
    \end{minipage}
    }
    \subfigure{
    \begin{minipage}{0.55\linewidth}
    \centering
    \includegraphics[width = \linewidth]{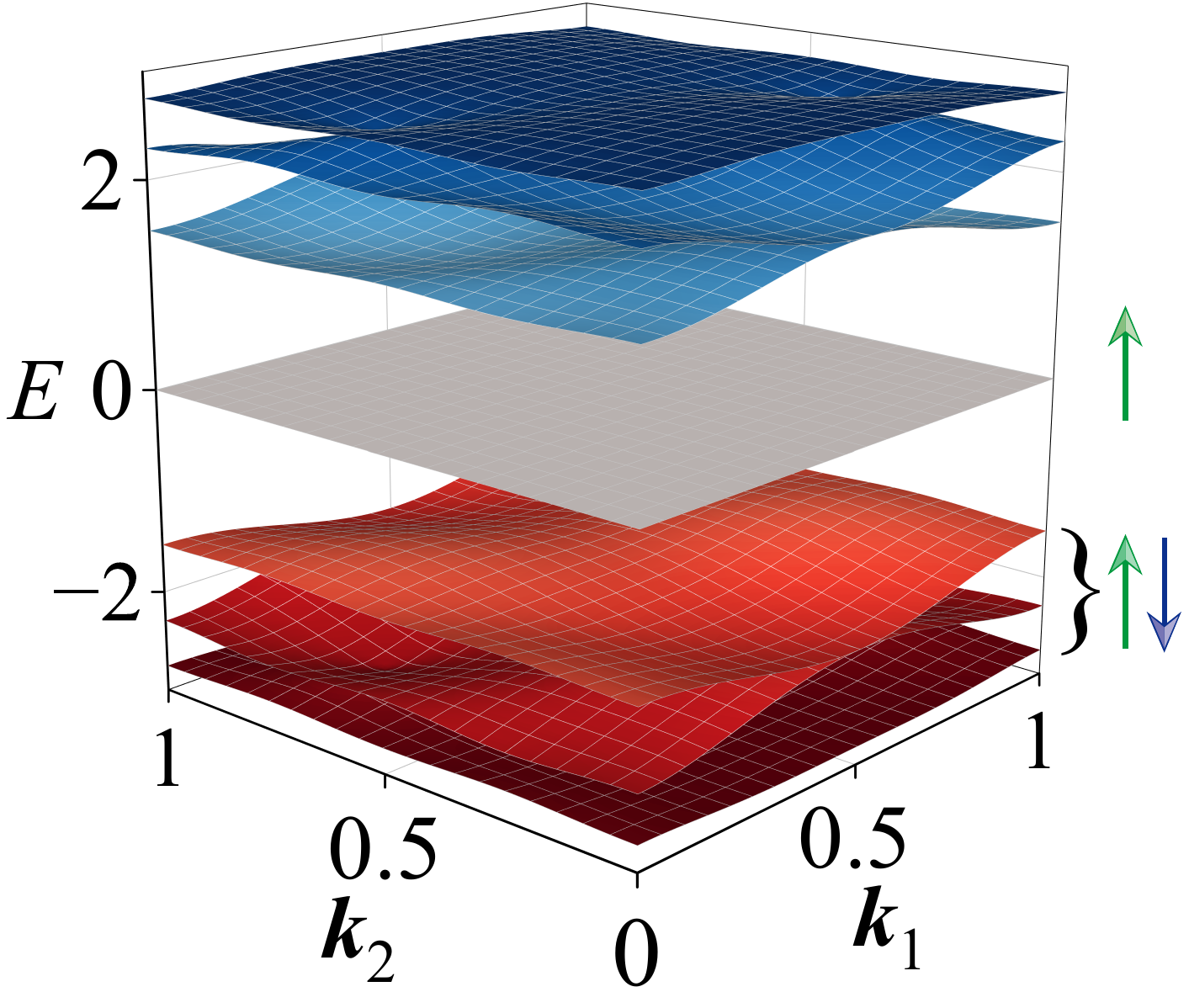}
    \put(-136,105){(d)}
    \label{fig:dispersion}
    \end{minipage}
    }%
    \subfigure{
    \begin{minipage}{0.38\linewidth}
    \centering
    \includegraphics[width = \linewidth]{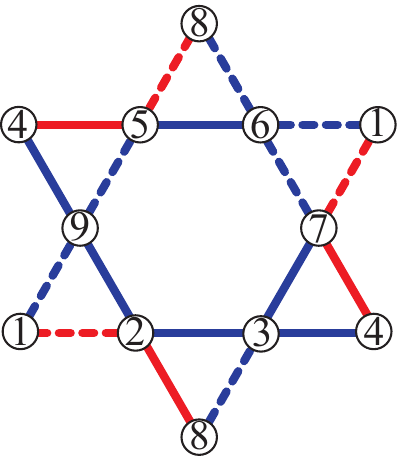}
    \put(-94,101){(e)}
    \label{fig:site_index}
    \end{minipage}
    }

    \caption{(a)-(c) depict the magnetization distributions $\mathcal{M}_{\mathrm{I}}$, $\mathcal{M}_{\mathrm{II}}$ and $\mathcal{M}_{\mathrm{III}}$ for the VBS-I,-II and -III states, respectively. The dashed black lines denote the $\sqrt{3} \times \sqrt{3}$ unit cells for these three states, while the solid yellow lines mark the corresponding positions of the David stars formed by hopping patterns, as shown in Fig.~\ref{fig:vbs_1}-\ref{fig:vbs_3}. (d) shows the spinon dispersion derived from the Ans\"atz of VBS-II state with optimal variational parameters in the $120 \times 120 \times 9$ system, where $\boldsymbol{k}_{1} = (3/2, \sqrt{3}/2)$ and $\boldsymbol{k}_{2} = (-3/2, \sqrt{3}/2)$ are the reciprocal vectors. The flat band with $E = 0$ is threefold degenerate. (e) labels the indices of the 9 sublattices in the VBS-II state.}
    \label{fig:mz_site_and_ek}
\end{figure}

Furthermore, we examine the magnetization distributions of the three degenerate VBS states, depicted in Fig.~\ref{fig:mz_site_1}, \ref{fig:mz_site_2} and \ref{fig:mz_site_3}, respectively. Remarkably, the magnetization moments exhibit fractional values $M_{i}/M_{s} = -1/3$ and $2/3$ within numerical error. To our knowledge, we do not find the similar example of a magnetic structure with fraction moments in a VBS state responsible for 1/3 magnetization plateau discussed before.
%Within numerical error, the magnetization values are limited to $M_{i}/M_{s} = -1/3$ and $2/3$.
Additionally, the magnetization patterns form a David-star shape and are related by the translational transformations $T_{1,2}$, such as $\mathcal{M}_{\mathrm{III}} \overset{T_{1}}{\longrightarrow} \mathcal{M}_{\mathrm{II}} \overset{T_{1}}{\longrightarrow} \mathcal{M}_{\mathrm{I}}$. The magnetization distributions obtained from the VMC calculations here differ significantly from that proposed in Ref.~\onlinecite{nat.commun.4.2287}, despite all of them exhibit David-star structures.

\begin{figure}
    \centering
    \subfigure{
    \begin{minipage}{\linewidth}
    \centering
    \includegraphics[width = \linewidth]{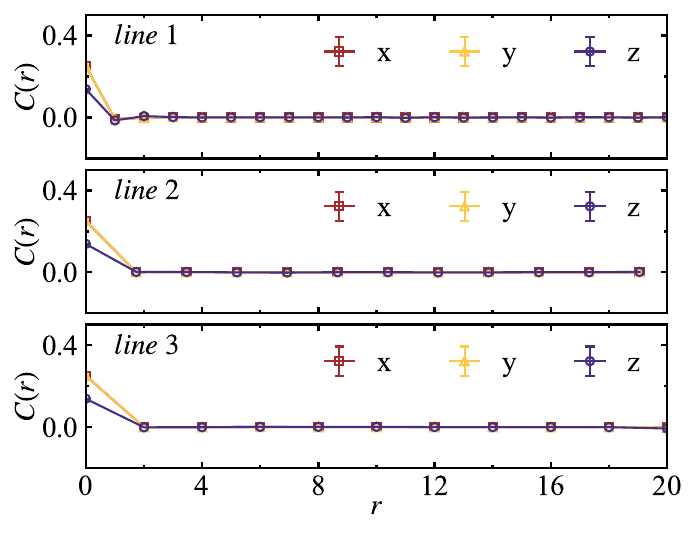}
    \put(-245,180){(a)}
    \label{fig:spin_cor}
    \end{minipage}
    }
    \subfigure{
    \begin{minipage}{\linewidth}
    \centering
    \includegraphics[width = \linewidth]{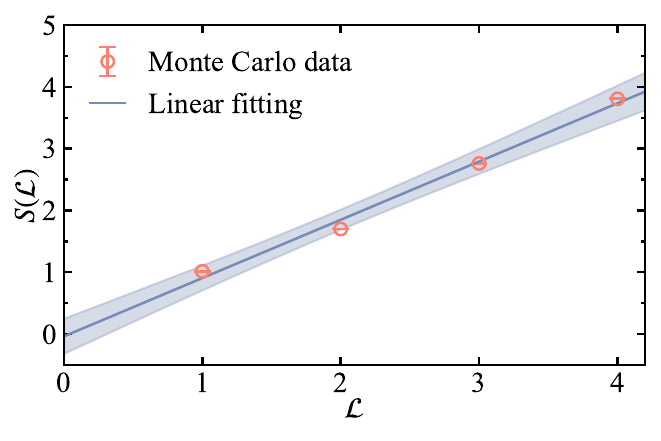}
    \put(-245,155){(b)}
    \label{fig:entropy}
    \end{minipage}
    }

    \caption{(a) Spin correlation $C(r)$ as a function of distance $r$ along the three green dashed lines in Fig.~\ref{fig:mz_site_2}, labeled with arrows $l_{1,2,3}$. The legend $\alpha \in \{x, y, z\}$ denotes the $\alpha$-component correlation $C(r) = \langle S_{0}^{\alpha} S_{r}^{\alpha}\rangle - \langle S_{0}^{\alpha} \rangle \langle S_{r}^{\alpha} \rangle$. (b) Entropy of the VBS-II state for a system size of $L_{1,2} = 12$. The best linear fit, $S(\mathcal{L}) = \alpha \mathcal{L} - \gamma$, yields a topological entanglement entropy $\gamma \approx 0.038 \pm 0.286$, where $\mathcal{L}$ represents the area of the subsystem as $\mathcal{L}^{2} \boldsymbol{a}_{1} \cdot \boldsymbol{a}_{2}$. The blue shading indicates the confidence interval for the linear fitting.}
    \label{fig:spin_cor_and_entropy}
\end{figure}

%We can also conduct further analysis based on the spinon dispersions.

Though the fractional spin moment distribution is believed to the origin from strong quantum fluctuations, we may still get some useful information from the spinon dispersion.
Figure~\ref{fig:dispersion} shows the spinon dispersion of the VBS-II state, which has the largest overlap with the wave function $|\Psi_{\mathrm{ED}} \rangle$, as listed in Tab.~\ref{tab:overlap_with_ed}. There are nine bands in total, where the spin-up (spin-down) spinons occupy six (three) bands accounting for the 1/3 magnetization. Notably, the spin-up spinons occupy three degenerate flat bands, which applies to all three degenerate VBS states.
%\widehat{}This indicates that there are three fully polarized spins per unit cell at the mean-field level.
To determine the contribution per site of the three flat bands, we define a weight factor as follows:
\begin{equation}
    f_{\alpha} = \sum_{n, \boldsymbol{k}} |\langle 0 |c_{\alpha \uparrow} \gamma_{n, \boldsymbol{k}}^{\dagger}|0\rangle,
\end{equation}
where $|0\rangle$ is the vacuum, $\alpha$ is the index of the sublattice as shown in Fig.~\ref{fig:site_index}, $\gamma_{n, \boldsymbol{k}}^{\dagger}$ is the creation operator of the mean-field quasi-particle with momentum $\boldsymbol{k}$ for the $n$-th flat band. We find two different weight factors: $f_{1,2,4,5,7,8} = 0.39377$ and $f_{3, 6, 9} = 0.21246$, indicating that the magnetization per site should take these values at the corresponding sublattices. Thus, the three flat bands are related to all nine sublattices rather than just any three, and the magnetic moment distribution structure is consistent with VMC calculations, disregarding the sign of the magnetic moment. Of course, the detailed distribution of magnetization is significantly influenced by many-body correlation effects, which deserve further investigation. We note that a similar argument applies to the VBS-I and -III states.

For the 1/3-magnetization plateau phase, the number of the sites in a unit cell is $n = 9$. Consequently, the relationship $n(S - M) = \frac{9}{2}(1 - \frac{1}{3}) = 3$ being an integer aligns with the theory of magnetization plateaus~\cite{nat.commun.4.2287, PhysRevLett.78.1984, PhysRevB.79.064412}. This implies all the three VBS states have no $\mathbb{Z}_{q} (q \ge 2)$ gauge structure. Conversely, for the $\mathbb{Z}_{3}$ QSL state at the 1/9-magnetization plateau~\cite{nat.commun.4.2287, PhysRevLett.133.096501}, $n(S - M) = \frac{3}{2}(1 - \frac{1}{9}) = 4/3$ due to the $\mathbb{Z}_{3}$ gauge field as mentioned in Ref.~\cite{PhysRevB.79.064412}. These results suggest that the denominator of $n(S - M)$  is closely related to the gauge structure of the many-body quantum state. This is an intriguing issue that warrants further study.

With the introduction of the finite field $B_{z}$, the system reduces the $SU(2)$ symmetry to $U(1)$, resulting in the transverse and longitudinal components of the spin correlation becoming non-equivalent. As shown in Fig.~\ref{fig:spin_cor}, we calculate all three components of the spin correlation $C(r)$ as a function of the distance $r$ between two sites along three different directions for the VBS-II state. The $x-$ and $y-$ are identical, differing from the $z$-component, which is consistent with $U(1)$ symmetry. Although the results indicate that the spin correlation is entirely short-range, similar to a QSL state, the topological entanglement entropy is nearly zero within numerical error, as shown in Fig.~\ref{fig:entropy}. This suggests that the VBS state lacks long-range entanglement and is topologically trivial according to the classification of topological order~\cite{nsr.1.68, RevModPhys.89.041004}.

{\it Conclusions.} In summary, we elaborate that the ground state of the 1/3 magnetization plateau phase in the spin-1/2 kagome antiferromagnetic Heisenberg model exhibits a triply degenerate $\sqrt{3} \times \sqrt{3}$ valence-bond-solid order, as revealed through variational Monte Carlo and exact diagonalization methods.
%We discover that this ground state is triply degenerate, characterized by three mutually orthogonal valence-bond-solid states, whose magnetizations are related by translational transformations.
These valence-bond-solid states exhibit fractional spin moments, which differ significantly from that previously proposed using the density matrix renormalization group method~\cite{nat.commun.4.2287} where fully polarized spin moments exist.
Additionally, while the spin correlations in these valence-bond-solid states are short-range, similar to quantum-spin-liquid states, they are topologically trivial, as indicated by the vanishing topological entanglement entropy.

\begin{acknowledgments}
This work was supported by National Key Projects for Research and Development of China (Grant No. 2021YFA1400400 and No. 2024YFA1408104) and the National Natural Science Foundation of China (No. 12434005, No. 12374137, and No. 92165205).
\end{acknowledgments}

\bibliography{ref}

\end{document}